\documentclass[10pt,twocolumn]{IEEEtran}
\usepackage{amsmath, graphics,amssymb,epsfig,subfigure,color,amsthm,cite}

\begin{document}
 
\title{{Advanced Interference Management Technique: Potentials and Limitations\\  } 
}

\author{\normalsize
  \IEEEauthorblockN{Namyoon Lee~\IEEEmembership{Member,~IEEE,} and Robert W. Heath Jr.~\IEEEmembership{Fellow,~IEEE} \thanks{N. Lee  is with Intel Labs, 
2200 Mission College Blvd, Santa Clara, CA 95054, USA (e-mail:namyoon.lee@intel.com). 
 R. W. Heath Jr. is with the Wireless Networking and Communications Group, Department of Electrical and Computer Engineering, The University of Texas at
Austin, Austin, TX 78712, USA. (e-mail:rheath@utexas.edu). This work was supported in part by the National Science Foundation under grant number NSF-CCF-1218338, and by a gift from Intel 5G program. }   } 
}
 
\maketitle

\begin{abstract}
Interference management has the potential to improve spectrum efficiency in current and next generation wireless systems (e.g. 3GPP LTE and IEEE 802.11). Recently, new paradigms for interference management have emerged to tackle interference in a general class of wireless networks: interference shaping and interference exploitation. Interference shaping is a technique that creates a particular linear combination of interference signals when transmitters propagate signals to minimize the aggregated interference effect at each receiver. Interference exploitation is a method that harnesses interference in decoding as side-information to improve data rates. Both approaches offer better performance in interference-limited communication regimes than traditionally thought possible. This article provides a high-level overview of several different interference shaping and exploitation techniques for single-hop, multi-hop, and multi-way network architectures with graphical illustrations. The article concludes with a discussion of practical challenges associated with adopting sophisticated interference management strategies in the future.  
\end{abstract}

%%%%%%%%%%%%%%%%%%%%%%%%%%%%%%%%%%%%%%%%%%%
\section{Introduction}\label{Section:Intro}
%%%%%%%%%%%%%%%%%%%%%%%%%%%%%%%%%%%%%%%%%%%

Interference is a fundamental phenomenon in wireless communication networks. It is a result of the superposition and broadcast nature of wireless transmission along with spectrum shared among multiple users. Uncoordinated interference reduces wireless network throughput. As a result, it is essential to understand and manage interference to achieve the highest network performance. 
 
Conventional approaches to deal with interference are 1) avoiding interference through orthogonalization of the shared time/frequency resource, 2) treating other transmitters' signals as noise, or 3) decoding interference. These strategies have been studied extensively and adapted to contemporary wireless systems such as cellular and wireless local area networks (WLANs). Although these approaches control interference without system overhead, it turns out that they are not optimal in most network configurations, except in certain special cases. For example consider the $K$-user interference channel where $K$ transmitters send data to their corresponding receivers in a shared wireless medium. When the interference power is of the order of the power of the signal of interest, for instance, the resource orthogonalization method has resulted in (at best) achieving the same data rate order as the rate of a single communication link because of its inefficient usage of the spectrum.

%When the interference power is of the order of the power of the signal of interest, the traditional interference management approaches have resulted in (at best) achieving the same data rate order as the rate of a single communication link because of their inefficient usage of the spectrum.

\begin{figure*}[t|]
  \centering
	\includegraphics[width=7.3in]{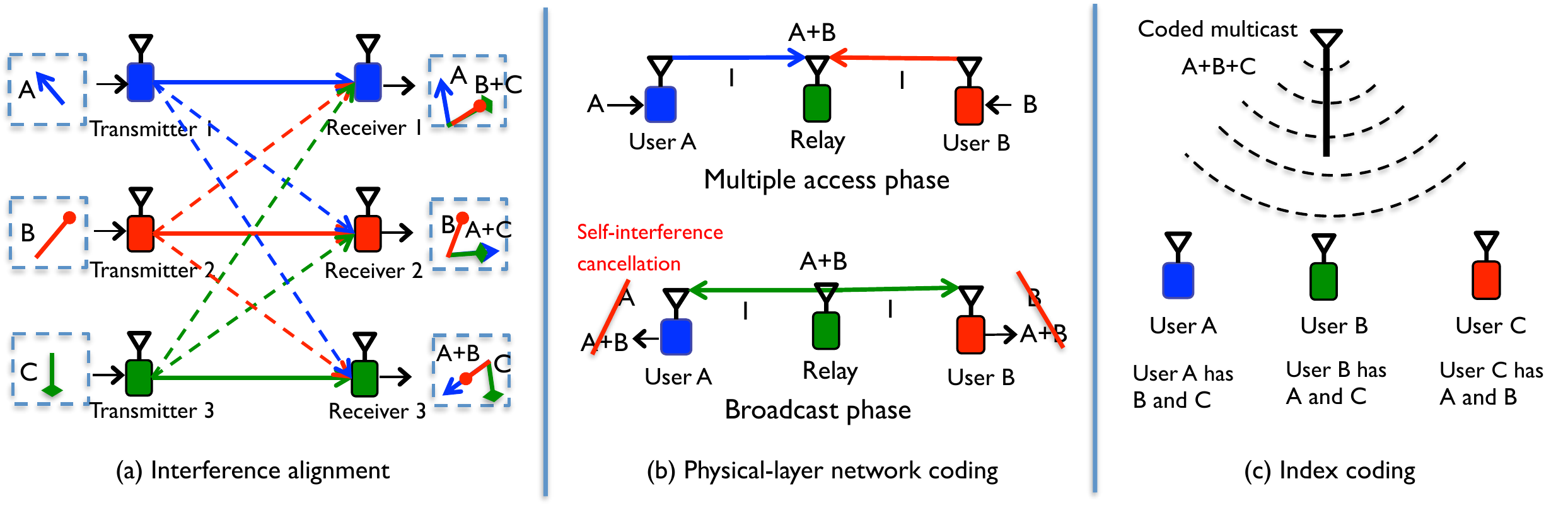}
	\caption{(a) An interference shaping strategy known as interference alignment at each receiver, three interferers collapse to appear as two. This enables interference free decoding in a desired signal subspace \cite{Jafar_Nowbook}. (b) An interference exploitation technique known as physical-layer network coding in a two-way relay channel. The two users exchange two packets within two time slots by using self-interference signal as side-information. (c) A coded multicast transmission strategy known as linear index coding in a broadcast channel. A transmitter sends a coded multicast signal to three users so that each user is able to decode a desired packet using knowlege of other users' packets as side-information.}
	\label{fig:1}
\vspace{8pt}
\end{figure*}

Recently, new paradigms for interference management strategies have emerged: interference shaping and interference exploitation. Both techniques offer better performance in the interference-limited communication regime than traditionally thought possible. The idea behind interference shaping is to create a certain interference pattern when transmitters propagate signals so that the aggregated interference effect is minimized or eliminated at each receiver. Interference alignment \cite{Jafar_Nowbook,MMK:08} and interference neutralization \cite{IN:08},  which is also known as distributed zero-forcing in \cite{Rankov_IN_2007}, are representative interference shaping techniques. The idea behind interference exploitation is to harness interference as a useful signal, in other words as side-information, to increase data rates. Network coding\cite{Katabi,Ahlswede_NC_2000,Rankov_IN_2007} and index coding \cite{Indexcoding_2006} are representative techniques for exploiting interference. A conceptual figure for interference alignment, physical-layer network coding, and index coding is illustrated in Fig. \ref{fig:1}, each of which will be explained with detail in Section II and IV.

In this article, we provide an overview of recent interference management strategies that leverage the ideas of interference shaping and exploitation from single-hop interference networks to multi-hop and multi-way interference networks with increasing levels of network complexity. We provide intuitive explanation for each strategy including figures to illustrate the main ideas without relying on the sophisticated mathematical expositions found in the original papers. Compared with previous overview papers \cite{Omar:12,Jeon:12}, we provide a more comprehensive overview of interference management techniques including interference alignment and neutralization as well as other interference management techniques. This article considers applications to single-hop, multi-hop, and multi-way networks. We conclude the article with a discussion about practical limitations faced by cutting edge interference management techniques in wireless systems and possible directions towards interference management research. For readers who are interested in the recent interference management techniques known as multicell/network multipoint transmission that are not covered in this article, we refer to \cite{Gesbert:2010}.

%%%%%%%%%%%%%%%%%%%%%%%%%%%%%%%%%%%%%%%%%%%
\section{Single-Hop Interference Networks}
%%%%%%%%%%%%%%%%%%%%%%%%%%%%%%%%%%%%%%%%%%%

Single-hop interference networks are good models for communication in cellular networks and wireless local area networks. The term single-hop means that the transmitter delivers its message directly to the receiver without using other nodes as a relay. Furthermore, every transmitters has unicast data for one or multiple destinations. Interference management in single-hop interference networks is complicated as the number of transmit and receive user pairs increases.  The difficulty arises because every transmitter needs to effectively control the transmitted signals' propagations towards the unattended receivers.

Interference alignment was among the first of a new class of interference management techniques \cite{Jafar_Nowbook} (please also refer to the references in \cite{Jafar_Nowbook} for further background). As depicted in Fig. \ref{fig:1}-(a), the idea of interference alignment is to align multiple interfering signals at each receiver to minimize the signal dimension occupied by the interference signals. As a result of this alignment, each receiver is able to decode its desired signal with no interference by projecting a received signal onto the null space of the interference signal space. Interference alignment has been proven to achieve the best possible sum of degrees of freedom (sum-DoF), also known as multiplexing gain, in a wide class of single-hop interference networks such as the interference channel and the X channel \cite{Jafar_Nowbook}. The sum-DoF is a coarse approximation of the sum-capacity of the networks, which describes how the sum of the rates scales with the log of SNR. Interference alignment makes it possible to linearly increase sum rates with the number of communication pairs in the idealistic case of an interference channel with time-varying channel coefficients \cite[Section 4.6]{Jafar_Nowbook}.

Although interference alignment makes it possible to provide  optimal throughput scaling in interference networks, it faces a number of challenges to be a practical interference management solution for many wireless systems \cite{Omar:12}. One problem is that the algorithms for implementing interference alignment over time-varying channels (also known as asymptotic interference alignment algorithms) may provide worse sum rate performance compared to that of a TDMA method in the finite SNR regime \cite[Section 4.6]{Jafar_Nowbook}. This comes from the fact that transmitters need to code over many independent channel realizations (i.e., needs for large channel diversity), which essentially sacrifices the desired signal power over multiple channel uses at the cost of asymptotic interference alignment. The cost is the signal dimension required to meet interference alignment conditions, which increases with the square of the number of transceiver pairs exponentially. To overcome this issue in time-varying channels, a near-optimal interference alignment algorithm called ergodic interference alignment was proposed to characterize an approximate ergodic capacity to within a constant number of bits/sec/Hz, regardless of SNR in certain interference networks (see the references in \cite[Section 4.5]{Jafar_Nowbook}). The crucial intuition behind ergodic interference alignment is the use of opportunistic transmission. Transmitters send signals using the carefully chosen two time indices that provide the complement channel coefficients. Each receiver is able to decode its desired signal with no interference by simply adding up the two received signals. The main drawback of ergodic interference alignment, however, lies in the impracticality in current wireless systems due to high delays and buffer constraints in hardware. 

\begin{figure*}[t|]
  \centering
	\includegraphics[width=6.3in]{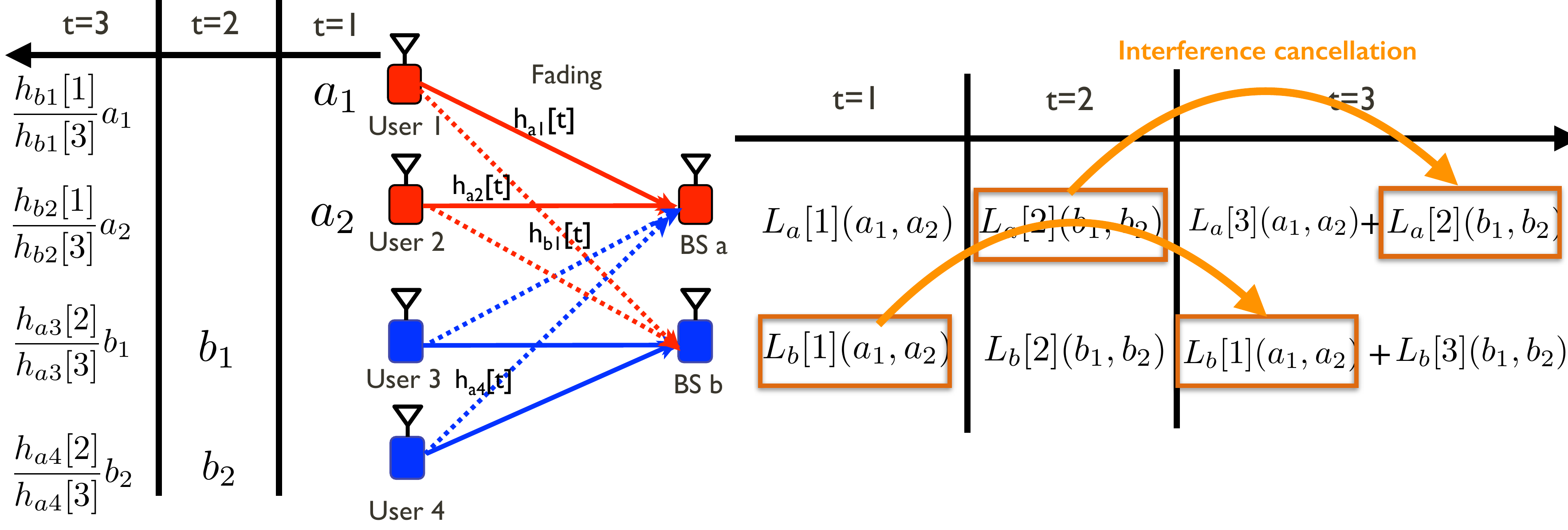}
	\caption{Diagrams illustrating space-time interference alignment in a two-cell two-user interfering multiple access channel. Each BS decodes two independent symbols from the two associated uplink users by spaning three time slots. }
	\label{fig:2}
\vspace{8pt}
\end{figure*}

Another direction for overcoming the performance loss in the finite SNR regime is to apply interference alignment algorithms using multiple signal dimensions inherently induced by multiple antenna systems \cite[Section 4.1]{Jafar_Nowbook} \cite{Omar:12,Bresler}. For example, for the multiple-input-multiple-output (MIMO) interference channel with a constant channel coefficient, assuming per-user power constraint, iterative interference alignment algorithms that maximize signal to interference plus noise ratio (Max-SINR) or minimize the sum of mean square errors (MMSE) were proposed in \cite[Section 4.1]{Jafar_Nowbook}, \cite{Omar:12}. By aligning interference signals distributively while maximizing the desired signal's strength, the iterative interference alignment algorithms  were found to achieve a higher total throughput compared to that of conventional interference management methods including TDMA in the finite SNR regime. One drawback of conventional interference alignment algorithms is that they do not scale the total throughput linearly with the number of communicating pairs due to limited channel diversity provided by the MIMO channels, provided that the number of antennas at each node is fixed. For example, for the symmetric $K$-user MIMO channel, each transmitter and receiver with $M$ antennas, it turned out that the maximum sum-DoF is at most $2M$ \cite{Bresler}. This  reveals a fundamental performance limitation of interference alignment for multi-antenna based wireless systems - the rates only scale with the number of antennas not the number of user pairs as in the time-varying case.

%In practice, however, the acquisition of such instantaneous and global CSIT as a means toward cooperation is highly challenging due to the distributed nature of transmitters and dynamic wireless propagation environments.

Another limitation for the interference alignment algorithms is that they require global and instantaneous channel state information at transmitter (CSIT), or overhead intensive over-the-air iterative design. To resolve this, recently, interference alignment has  moved in the direction of using limited channel knowledge to make the benefits of alignment more readily available in practical wireless systems. Blind interference alignment (see the references therein \cite[Section 4.9]{Jafar_Nowbook}) is a class of interference alignment techniques that use knowledge about autocorrelation functions of the channels instead of channel values themselves, thereby requiring less CSIT. The idea is that a transmitter carefully and repeatedly sends symbols based on knowledge of the channel correlation so that each receiver sees the same aggregated interference. This algorithm leads a receiver to simply subtract the aggregated interference and then resolve the multiple desired signals free from interference.  Although blind interference alignment exhibits a possibility of exploiting interference alignment under limited CSIT, it is  difficult to implement in current wireless systems due to special requirements such as the staggered block fading structure or the receive antenna switching method with reconfigurable antennas. Assuming a general fading scenario where channels independently vary over time or frequency, an innovative transmission strategy called MAT method \cite{Maddah-Ali:12} using limited CSIT was proposed in the context of the vector broadcast channel. The core idea was that a transmitter sagaciously uses completely-delayed CSIT (i.e., CSI feedback delay larger than the channel coherence time) to align inter-user interference between the past and the currently observed signals. This method, surprisingly, revealed that completely delayed CSIT is still useful in providing substantial DoF gains compared to no CSIT. The essence of the MAT scheme was extended in interference networks by devising retrospective interference alignment \cite[Section 4.10]{Jafar_Nowbook}. This approach showed that completely delayed CSIT also leads to the DoF improvement even in interference networks consisting of distributed transmitters.

 \begin{figure*}[t|]
  \centering
	\includegraphics[width=5.3in]{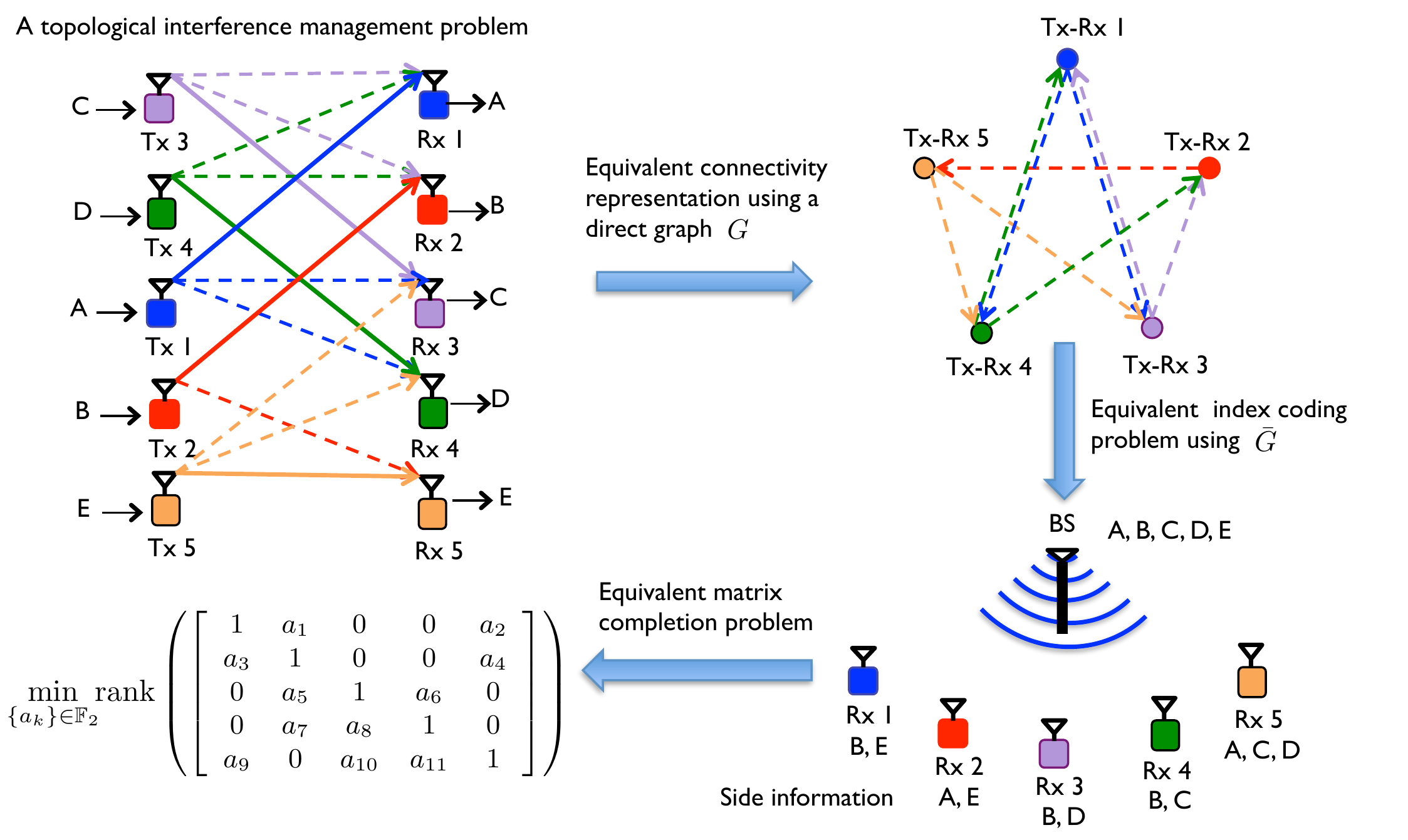}
	\caption{Diagrams illustrating the connection among a topological interference management problem, the equievalent linear index coding problem, and the matrix completion problem. The left-top figure illustrates a partially connected interference channel where the solid and dotted lines represent the desired links and interfering links, respectively. The connectivity of interfering links is equivalently represented using a direct graph. Using the complement of the graph as a side-information graph, the equivalent linear index coding problem is illustrated as shown in the lower-right figure. This linear index coding problem can also be formulated to a matrix completion problem over a binary field as depicted in the lower-left figure.}
	\label{fig:3}
\vspace{8pt}
\end{figure*}

The idea of blind interference alignment was advanced in different directions based on the limited CSIT, including moderately delayed CSIT (feedback delay is less than the channel coherence time) \cite{Namyoon_STIA}, mixed CSIT (perfect delayed and partial instantaneous), and alternating CSIT \cite{Tandon:13}. In particular, under the moderately delayed CSIT model, space-time interference alignment in \cite{Namyoon_STIA} was found to achieve the optimal sum-DoF is attainable for a certain class of interference channels even if the feedback delay exists, provided that  the feedback delay is less than a certain fraction of the channel coherence time. The idea of space-time interference alignment is illustrated in Fig. \ref{fig:2} for a two-cell multi-user uplink communication scenario where two users placed at the cell-edge area wish to communicate with their respective base station (BS) by sharing the same time-frequency resource. Under the premise of the moderately-delayed CSIT model, each user employs (perfect) outdated CSIT for the two-thirds of the channel coherence time, while it exploits (perfect) current CSIT for the remaining fraction of it. In such a setting, it is possible to show that the four users are able to send four independent information symbols $a_1$, $a_2$, $b_1$, and $b_2$ to the two BSs by spanning three time slots $t_1$, $t_2$, and $t_3$ that are belong to different channel blocks. In time slot one, the two users in cell 1 send their information symbols $a_1$ and $a_2$. Then, BS 1 and BS 2 obtain the linear combinations of the two symbols, which are denoted by $L_a[1](a_1,a_2)$ and $L_b[1](a_1,a_2)$, where the linear coefficients are chosen from the channel values between the transmitter and receiver. In the second time slot, the two users in cell 2 transmit information symbols $b_1$ and $b_2$. Then, similar to the first time slot, each BS receives a linear combination of $b_1$ and $b_2$, $L_a[2](b_1,b_2)$ and $L_b[2](b_1,b_2)$. In the third time slot, the users exploit the current and delayed CSIT jointly so that the two BSs receive the same interference equations that previously received. To accomplish this, each user applies a precoding coefficient that inverts the current channel response from the user to the interfering BS and, then multiplies the same channel coefficients that the interfering BS observed previously. This interference alignment in the third time slot allows for each BS to perform successive interference cancellation with the overheard interference equation during the first and second time slots; thereby, the desired information symbols are reliably decodable at each BS.

Following upon the success of the interference alignment techniques with limited CSIT, more recent and emerging work has considered developing interference management techniques using extremely less knowledge of CSIT. Topological interference management in \cite{Jafar_Index_coding} is an ingenious approach in that direction. The idea is to smartly schedule multiple transmissions based on network connectivity information at transmitter. This approach was shown to achieve  considerable DoF gains for a certain class of partially connected interference channels only with network connectivity information. Remarkably, topological interference management has also received attention because of its connections to linear index coding \cite{Indexcoding_2006}. As illustrated in Fig. \ref{fig:3}, a partially connected interference channel can be represented by a direct graph $G$ where each vertex corresponds to a transmit-and-receive pair and the direct edges represent the connectivity of interfering links. By taking the complement of the direct graph, $\bar{G}$, it was shown   \cite{Jafar_Index_coding} that topological interference management can be converted into a linear index coding problem with side-information graph $\bar{G}$. 

\begin{figure*}[t|]
  \centering
	\includegraphics[width=6.3in]{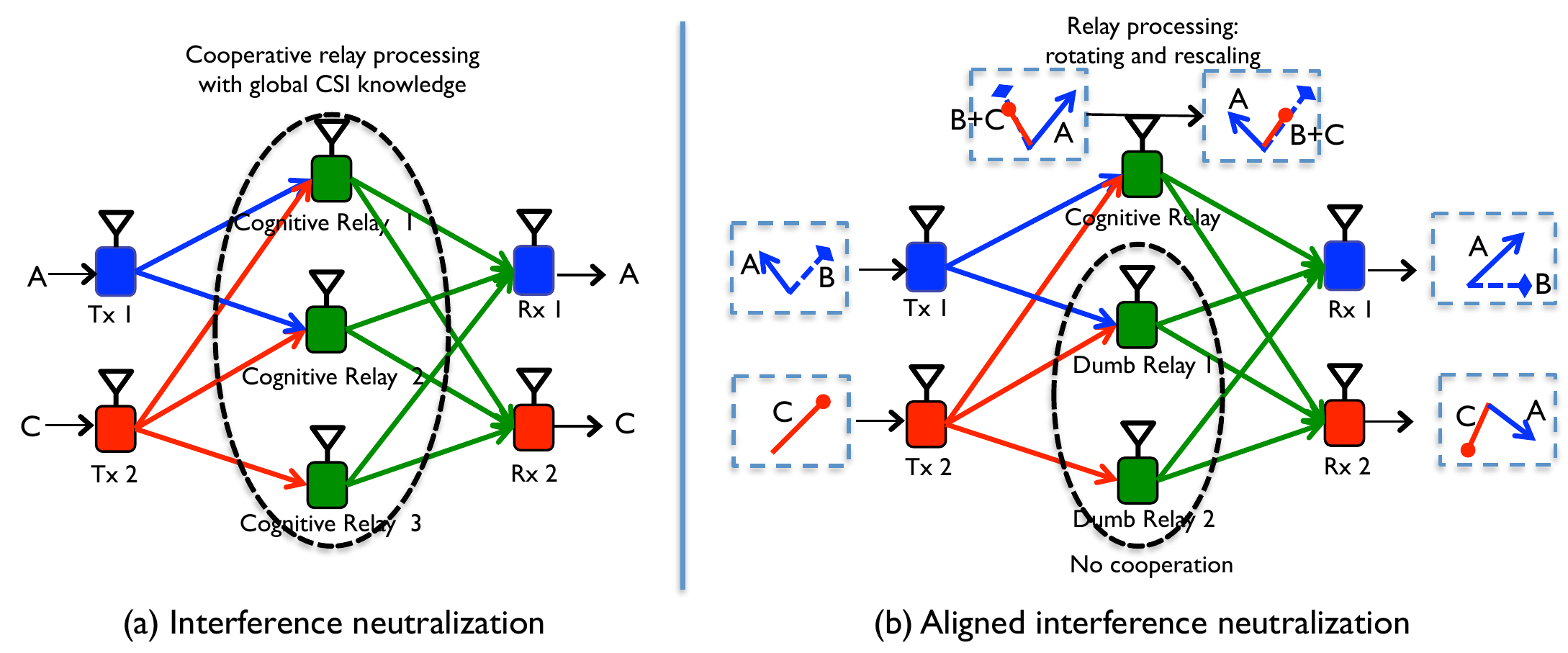}
	\caption{(a) A diagram illustrating interference neutralization (distributed zero-forcing) in a two-pair two-hop interference channel with three relays. At each destination, no interference appears by carefully choosing relaying strategies. This enables interference free decoding for a desired signal. (b) A diagram illustrating aligned interference neutralization in a two-pair two-hop interference channel with a cognitive relay and two dumb relays.}
	\label{fig:4}
\vspace{8pt}
\end{figure*}

Finding the optimal solution for the topological interference management problem is challenging for the $K$-user interference channel with an arbitrary network connectivity. Using the strong relation to linear index coding, however, one can find the upper and lower bound on the symmetric DoF for the topological interference management problem with an arbitrary network connectivity. This is because, for the given direct side-information graph $\bar{G}$, the optimal linear index code length, $L$, is lower and upper bounded by the maximum independent set number of the corresponding graph, $\bar{G}$, and the chromatic number of its complement, $G$. These graph-theoretical approaches also can be used to characterize the bounds on the symmetric DoF, $\frac{K}{L}$ for the corresponding topological interference management problem. Algebraic approaches can also be effective methods to solve topological interference management problems. Using the fact that the optimal linear index code length is equal to the minimum rank of a matrix that fits the side-information graph $\bar{G}$, it is possible to solve topological interference management problems by equivalently solving a matrix completion problem over a finite field as illustrated in Fig. \ref{fig:3}. Despite the strong relationship to linear index coding, finding the optimal solutions for the topological interference management problems via linear index coding is NP-hard for a general network connectivity. It would be a good direction to design an algorithm that solves any topological interference management problems with polynomial time complexity, while providing a solution close to the optimal symmetric DoF.

%%%%%%%%%%%%%%%%%%%%%%%%%%%%%%%%%%%%%%%%%%%
\section{Multi-Hop Interference Networks}
%%%%%%%%%%%%%%%%%%%%%%%%%%%%%%%%%%%%%%%%%%%

Multi-hop communication, in this article, refers to the case where a source and destination communicate through one or more other nodes that act as a relay. As in the single-hop case, we consider single-way configurations where a source only talks with a single destination. Interference management in multi-hop networks is sophisticated as relay nodes between the source-destination pairs propagate not only the desired signals  but also interference signals on the network. This makes it difficult to design relay strategies because it is ambiguous as to what extent a relay should forward, remove, align, or otherwise manage interference.

Interference management problems for multi-hop interference channels have been tackled by considering different network connectivity configurations. One popular network connectivity model is a layered two-hop interference channel where there is no direct channel paths between transmit and receive pairs; thereby relays are indispensable for the communication. Interference neutralization was proposed in \cite{Rankov_IN_2007} to overcome the interference barrier for such a layered multi-hop interference channels. The main idea of interference neutralization is a careful selection forwarding strategies at relays that results in the cancellation of interference signals through multiple channel paths. As a result of neutralization, each destination node is capable of decoding the desired signal without interference. By employing interference neutralization, it was demonstrated that interference-free communication is possible for a certain class of two-hop interference networks, provided that the number of relays is sufficiently larger than the number of source and designation pairs \cite{Rankov_IN_2007}. Fig. \ref{fig:4}-(a) illustrates an example of interference neutralization in a two-pair two-hop interference channel. The core idea of interference neutralization is the design of the relay precoder so that the sum of effective interference channel coefficients become zero at each receiver. For example, in the first hop, transmitter 1 and 2 send signals $A$ and $B$ and each relay receives a linear combination of $A$ and $B$. In the second hop, each relay applies the relay processing coefficient that rotates and rescales the mixed signal of $A$ and $B$. Since there are two interference neutralization equations with three variables, one can find an infinite number of solutions for the relaying coefficients. Accordingly, each receiver is able to perform interference-free decoding as the aggregated interference becomes zero at each receiver. For a general two-hop (layered) interference network comprised of $K$ source-destination pairs and $L$ relays, the number of required relays for interference neutralization is $L\geq K(K-1)+1$.

  \begin{figure*}[t|]
  \centering
	\includegraphics[width=6.6in]{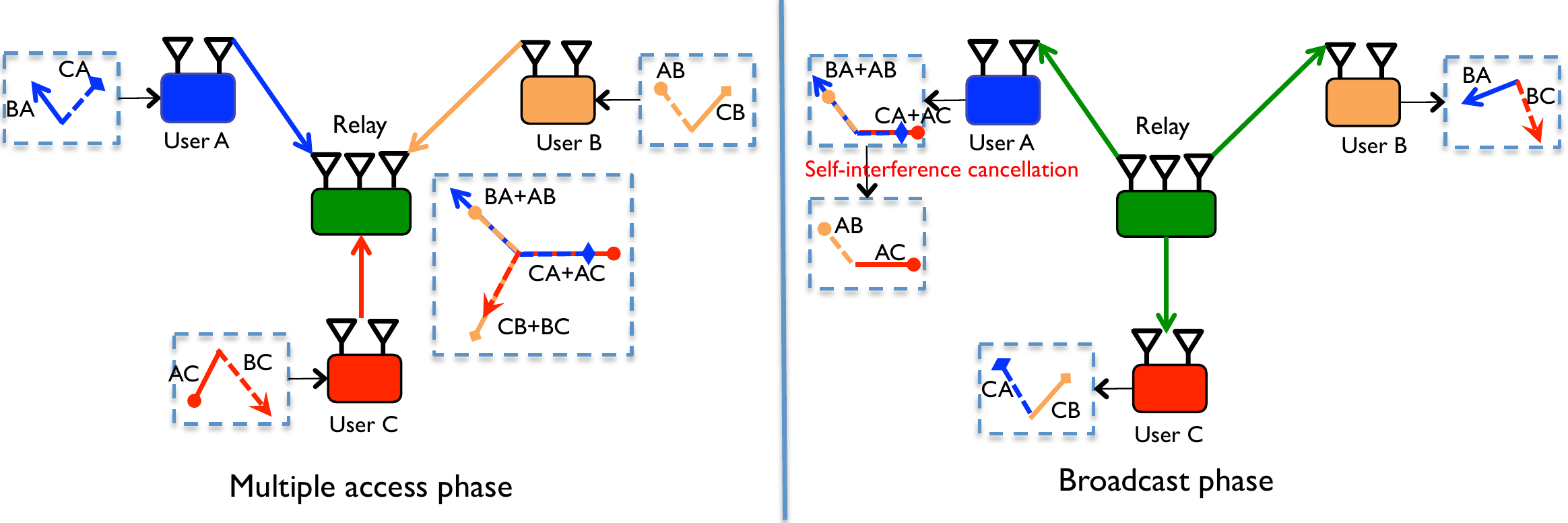}
	\caption{A diagram illustrating signal space alignment for network coding in a MIMO Y channel.}
	\label{fig:5}
\vspace{8pt}
\end{figure*}

The idea of interference neutralization was advanced by leveraging the idea of interference alignment. Considering precoding techniques over multiple symbols, aligned interference at relays can be neutralized through multiple paths; this is referred to as aligned interference neutralization \cite{Gou}, \cite{namyoon2}. Using the idea of neutralizing interference signals after aligning them at the relays, it was demonstrated that the optimal sum DoF of the two-pair two-hop interference network is achievable with two cooperative relays \cite{Gou}. The core idea of aligned interference neutralization in \cite{namyoon2} is illustrated in Fig. \ref{fig:4}-(b). In this channel, a cognitive relay has global channel knowledge while two dumb relays simply amplify and forward their received signals with no knowledge of channel conditions. In the first hop, transmitter 1 sends two signals $A$ and $B$ while transmitter 2 transmits signal $C$ over two time slots  using CSI knowledge between the transmitters and the cognitive relay so that the cognitive relay sees the aligned signal $B+C$ in a signal dimension while observing $A$ in a separate signal dimension. By carefully choosing the relaying coefficients over the two signal dimensions, it is possible to neutralize the aggregated interference propagated from the dumb relays and the cognitive relay at both receivers, which allows for the two receivers to extract their desired signals. Recent work \cite{Shomorony} for two-hop interference networks with $K$ transmitters, $K$ relays, and $K$ receivers showed that the cut-set bound (i.e., sum-DoF=$K$) is achievable asymptotically when the number of channel diversity large enough. The key idea in showing this result was a variant of aligned interference neutralization, which diagonalizes the two-hop network with distributed relays by canceling the interference between all transceiver pairs.

Another popular network connectivity model is a non-layered multi-hop interference channel where there are direct channel paths among all transmit and receive pairs including relays in the channel. Under this fully-connected multi-hop channel model, unlike the layered multi-hop interference channel, it was demonstrated that the use of relays cannot increase sum-DoF of the channel \cite{JafarRelay}. A major beneficial aspect of exploiting relays in the non-layered multi-hop interference channel, however, was found by recent work \cite{Tian}. The benefit is the considerable reduction of CSI feedback amount required for interference alignment compared to the method in \cite{JafarRelay}. By proposing a relay-aided blind interference alignment technique, it was shown that the optimal sum-DoF of the channel is achievable without CSIT, provided that the relay with global CSI forwards the received signals to receivers in such a way that aligns interference. Similar to blind interference alignment \cite[Section 4.9]{Jafar_Nowbook}, this interference management technique exhibits a possibility to mitigate interference successfully with limited CSIT in the non-layered multi-hop interference channel.

%%%%%%%%%%%%%%%%%%%%%%%%%%%%%%%%%%%%%%%%%%%
\section{Multi-Way Communication Networks with Relays} 
%%%%%%%%%%%%%%%%%%%%%%%%%%%%%%%%%%%%%%%%%%%

Multi-way communication networks with relays models a general communication scenario where multiple users exchange information with the help of multiple relay nodes. This wireless network architecture has received attention because of its  broad applications to cellular networks, sensor networks, and D2D communication. The most distinctive feature of multi-way relay networks are that source and destination nodes are inseparable. This feature introduces a new potential of exploiting interference signals as an efficient interference management approach.

Network coding and index coding are representative techniques for exploiting interference. For multi-hop wired networks, \cite{Ahlswede_NC_2000} originally introduced the idea of network coding to boost network throughput. The crucial principle underlying network coding is that intermediate nodes in networks forward functions (a linear combination) of their received packets, instead of independent packets. The same spirit of exploiting interference was extended into wireless networks. In the two-way relay channel in which two nodes exchange information with each other via a shared relay node, the concept of physical layer (analog) network coding \cite{Katabi,Rankov_IN_2007} was introduced. As illustrated in Fig. \ref{fig:1}-(b), physical-layer network coding allows the exchange of packets $A$ and $B$ within two communication phases by canceling the known self- interference signal a prior. Specifically, in the first phase, the two users send packet A and $B$ to the relay simultaneously. In the second phase, the relay broadcast the linear combination of $A$ and $B$ to the two users. Since each user has knowledge about self-interference signal a prior, it can cancel the self-interference effect from the received signal. This leads the performance improvement double compared to the case of the resource orthogonalization method. Drawing upon this idea, \cite{Rankov_IN_2007} showed that the rates of information exchange are double of these achieved by  traditional interference management strategies. 

The idea of linear index coding was initially introduced in \cite{Indexcoding_2006} for a noiseless broadcasting channel, in which a transmitter sends packets for multiple users, each with other users' packet in prior as side-information. The transmitter desires to deliver multiple packets to their respective users over a shared noiseless link. The principal goal of index coding is the design of transmit codes that minimize the number of required channel uses, while making all users decode the desired packets with the received signals from the transmitter and their own side-information. As depicted in Fig. \ref{fig:1}-(c), for instance, a transmitter desires to deliver packet $A$, $B$, and $C$ to user A, B, and C, respectively, and each user has knowledge of the others' desired packets. In this case, the optimal transmission strategy is to broadcast one XORed packet $A+B+C$ to all three users by spanning a channel use, as each user decodes the desired packet using the received $A+B+C$ and stored side-information $\{B,C\}$, $\{A,C\}$, and $\{A,B\}$.

%\subsubsection{Signal Space Alignment for Network Coding}

The concept of interference exploitation was extended by  combining the ideas of interference alignment and network coding; this is referred to as signal space alignment for network coding \cite{Lee_Lim_Chun:10}. Consider a three user multi-way communication scenario where a user with two antennas wants to exchange two independent messages each via a relay with three antennas; this is referred to as a MIMO Y channel \cite{Lee_Lim_Chun:10}. As illustrated in Fig. \ref{fig:5}, when each user sends two independent messages to the relay in the multiple access phase, the relay observes a total of six incoming signal streams. Since a relay is able to use a three dimensional signal space, the users cooperatively send the signals so that a pair of two signals for network coding can each be aligned as a signal dimension. In the broadcast phase, the relay applies the multi-user beam forming technique that cancels unmanageable interference to the users. At the user end, each user is able to first subtract its own transmit signal, the so-called self-interference, and then extract the desired signal from its partner, in a process that is analogous to network coding. This approach achieves the optimal sum DoF in the MIMO Y channel and in a broad class of multi-way communication networks with relays. 

\begin{figure}[t|]
  \centering
	\includegraphics[width=3.6in]{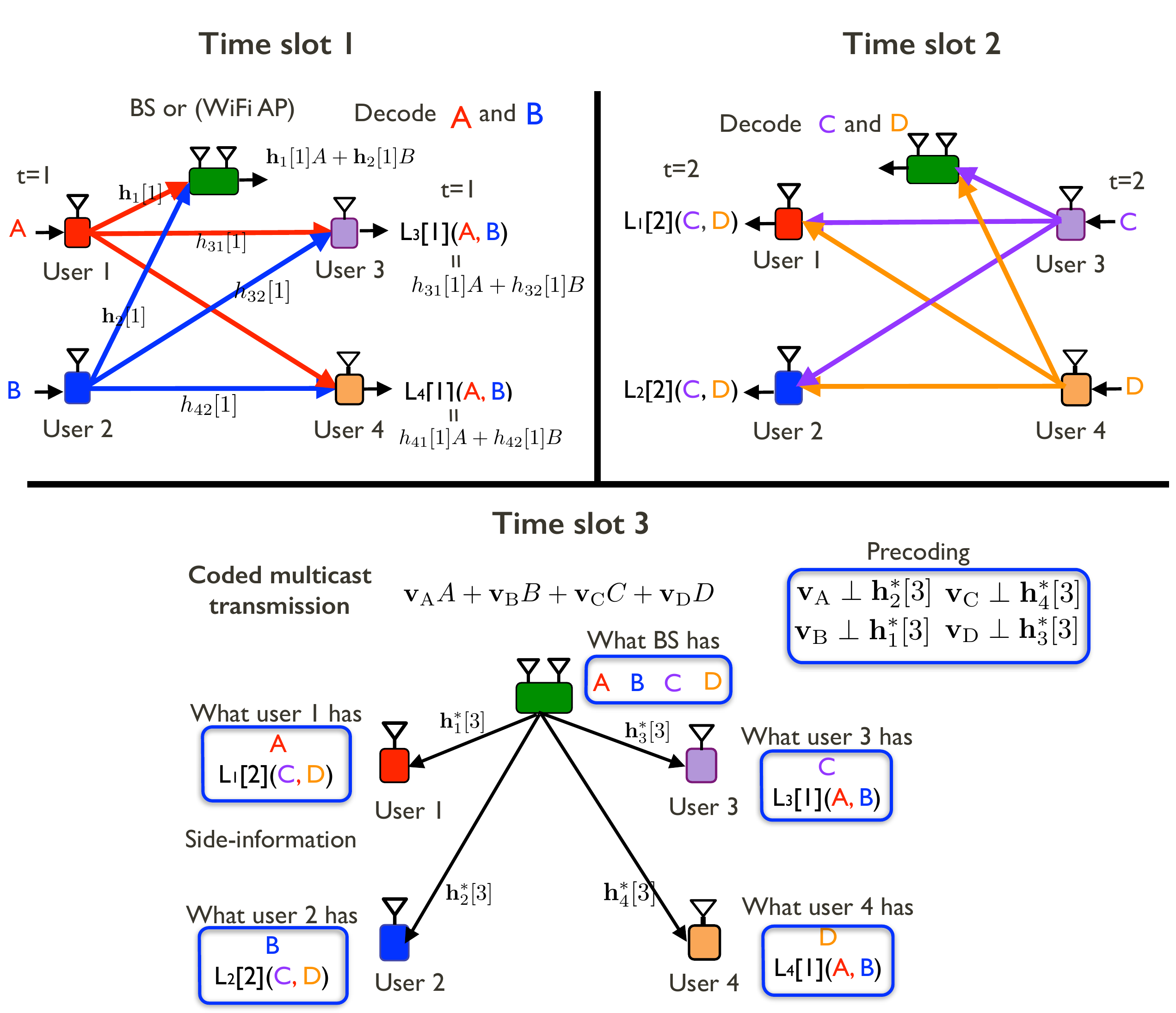}
	\caption{A diagram illustrating space-time physical-layer network coding in a two-pair two-way interference channel with a two-antenna relay.}
	\label{fig:6}
\vspace{8pt}
\end{figure}

Recently, the interference exploitation idea also was applied to fully-connected multi-way relay channels. Unlike the prior physical-layer network coding approaches that only exploit the self-interference signal as the main source of side-information, a new physical-layer network coding strategy called \textit{space-time physical layer network coding} was proposed in \cite{Namyoon_STPNC}. The essence is the exploitation of overheard interference signals as side-information in addition to self-interference signals for fully-connected multi-way relay networks. As illustrated in Fig. \ref{fig:6}, consider a fully-connected multi-way information exchange scenario where two pairs (user 1-3 and user 2-4) exchange packets with their partners via a relay with two antennas. This scenario can model the case where two D2D user pairs cooperatively exchange video files by sharing a WiFi access point. In time slot 1, user 1 and user 3 send two packet $A$ and $C$. Then, since user 2 and user 4 have a single antenna, the both cannot decode the desired packet reliably due to mutual interference. Instead, they just store the overheard linear combination of two packets $A$ and $C$ to exploit it later as side information. Whereas, the access point with two antennas is able to decode the two packets reliably by applying a zero-forcing decoder. In time slot 2, user 2 and 4 send packets $B$ and $D$. Then, similarly, user 1 and user 3 store the linear combination of $B$ and $D$, while the access point decodes the two packets thanks to the multiple antennas. Recall that during the previous two time slots, the access point has decoded all the packets, $A$, $B$, $C$, and $D$. Furthermore, each user has obtained two different types of side-information: 1) what it sent and 2) what it overheard. In time slot 3, the access point sends out a linear combination of the four packets using space-time precoding so that every users exploits both different types of side-information simultaneously. For example, the precoding vector carrying packet $B$ is selected so that the packet $B$ does  not propagate to user 1 who does not have knowledge of packet $B$ as side information. By carefully choosing precoding vectors with the same principle, every users is able to decode the desired packets successfully from the received signal in time slot 3 using the two types of side information obtained in the prior time slots. 
 
Space-time physical layer network coding can be explained through the lens of linear index coding in a general fully-connected multi-way relay network configuration comprised of a single antenna $K$ users and a relay equipped with $M$ antennas. The information exchange among users takes place using two phases: i) side-information learning and ii) space-time relay transmission. In the first phase, using the fact that the relay is able to decode $M$ independent packets per channel use thanks to multiple antennas, $M$ users among $K$ users are scheduled to send packets into the network at each channel use until the relay acquires full knowledge of packets in the network. During this phase, the remaining $K - M$ users overhear the linear combinations of transmitted packets per channel use, thereby learning the interference patterns. In the second phase, the relay starts to control information flows by broadcasting the superposition of packets so that all the users harness their side-information in the form of: 1) what they sent and 2) what they overheard in decoding. See the details for this generalization in \cite{Namyoon_STPNC}.

%\subsubsection{Index Coding}
%\begin{figure}[t|]
%  \centering
%	\includegraphics[width=4.3in]{Fig/Fig_Section1_IC.pdf}
%	\caption{A diagram illustrating index coding.}
%	\label{fig:IC_concept}
%\vspace{8pt}
%\end{figure}

%%%%%%%%%%%%%%%%%%%%%%%%%%%%%%%%%%%%%%%%%%%
\section{Research Challenges }
%%%%%%%%%%%%%%%%%%%%%%%%%%%%%%%%%%%%%%%%%%%

Despite the theoretical performance gains of the advanced interference management techniques described in the previous sections, there are many research challenges that remain to transform theory to practice. 
 
\subsection{Out-of-Cell Interference}

Most of the interference management techniques reviewed in this article were developed in idealized interference network settings. These settings consider a particular number of cooperative transmit-and-receiver pairs, ignoring the potential interference from outside of the cooperative set and often deemphasizing noise effects. There exists apparent limitations in translating the performance gains obtained from the advanced interference management techniques into practical wireless systems due to their simplified natures.

The practical challenge is to reevaluate the gains obtained from the interference management algorithms using models that accurately capture the impact of the irregular spatial structure of wireless node locations and channel propagation characteristics depending on operating frequency bands (e.g., low frequency or mmWave bands). One possible direction is to use analytical models for large-scale interference networks via stochastic geometry. This approach facilitates compact expressions of coverage probability and spatially averaged spectral efficiency of in networks with the effects of out-of-cell interference and channel propagation. For example, by adopting the downlink cellular network model via stochastic geometry \cite{Andrews_Tractable}, it would be interesting in future work to analyze the system-level performance of the recent interference management algorithms. This approach can gauge the values of the interference management algorithms with the eyes of system-level-performance.

\subsection{Partial CSIT}
A major problem found in most of the advanced strategies is that they substantially increase transmission rate only when ample knowledge of instantaneous and global CSIT is available. For instance, interference alignment and neutralization need global and perfect knowledge of CSIT across networks. In practice, however, the acquisition of such instantaneous and global CSIT as a means toward cooperation is highly challenging due to the distributed nature of transmitters and dynamic wireless propagation environments. In many limited CSIT scenarios, the promising gains from interference management strategies using instantaneous and global CSIT disappear, often providing the same result as cases where there is no CSIT. To resolve this issue, recently, interesting interference alignment algorithms  \cite{Namyoon_DSTIA, Gesbert2013} were proposed by using incomplete CSIT or local CSIT for feedback overhead reductions.

 Continuing the same spirit, to achieve potential gains in practical wireless systems, an interesting direction for future study is to explore the effects of using only statistics of the CSIT. For example, one may devise an efficient interference management technique that only requires the mean or covariance of the channel at transmitter, i.e., the first-order or second-order statistics of the channel. This approach could possibly reduces the amount of channel feedback significantly and increases the robustness for the CSIT uncertainty by delay and CSI quantization, at the expense of reduced performance compared with perfect and complete CSIT. 

\subsection{Link Scheduling through Index Coding}
%As mentioned in Section II, link scheduling algorithms based on index coding are promising as they only require to know information about network connectivity. 

Designing link scheduling algorithms based on index coding are promising as they only require to know information about network connectivity, causing the minimal CSIT acquisition overhead in wireless systems.  One major challenge in realizing the link scheduling algorithms in practical wireless systems is to design control signals that allows to acquire global link connectivity information at transmitters with extremely low signaling overheads. 

From an algorithm development side, many interesting problems still remain. One possible research direction is to explore the link scheduling algorithms using index coding when each wireless device has multiple antennas. Since most of current wireless systems employ multi-antenna transmission techniques, this approach would bring the additional performance improvements through multiplexing gains offered by MIMO systems. Another possible research direction is to develop interference management techniques using link connectivity information in multi-hop interference networks. For example, by exploiting network connectivity information of multi-hops jointly rather than employing network connectivity information per hop separately, it would be interesting to devise the joint scheduling algorithms among source and relay nodes in multi-hop interference channels. This approach possibly opens a new approach that manages interference with very limited channel knowledge in multi-hop interference networks.
 
% it would be interesting to investigate the performance of link scheduling algorithms from an ergodic perspective, considering time-varying network connectivity. 

\subsection{Interference Shaping in mmWave Systems}
Millimeter wave (mmWave) carrier frequencies promising candidate for next-generation cellular and WLAN (802.11ad and 802.11ay) wireless networks. MmWave offers the potential to support gigabit-per-second data rates due to both the vast bandwidth available in mmWave bands and the use of the large number of antenna arrays packed in a small area. Despite the gains, there are many practical challenges in the realization of mmWave wireless systems. One of the major challenges is the use of a hybrid MIMO processor that consists of a radio frequency chain (analog) beamformer and a baseband MIMO (digital) beamformer jointly to compensate excessive pathloss in mmWave bands while obtaining MIMO gains. This new hybrid physical-layer aspect changes the architectures of the channel estimation and feedback, which brings new challenges of using the advanced interference management methods. Therefore, it would be interesting to explore interference alignment or neutralization algorithms for multiuser transmissions under the hybrid MIMO processor architecture for mmWave systems.

%%%%%%%%%%%%%%%%%%%%%%%%%%%%%%%%%%%%%%%%%%%
\section{Conclusion}
%%%%%%%%%%%%%%%%%%%%%%%%%%%%%%%%%%%%%%%%%%%
Interference management is important for every wireless network. This article provided a  high level introduction to several recently developed interference management strategies. These strategies use the principles of interference shaping and interference exploitation to achieve better performance compared with systems that neglect interference. The potential gains and limitations for the new interference management techniques were discussed in various interference network scenarios including single-hop, multi-hop, and multi-way. The article concluded with a discussion of some relevant research directions that make interference management  more practical. The frontier for practical interference management techniques remains vast.

\end{document}